\documentclass{article}

\usepackage{arxiv}

\usepackage[utf8]{inputenc} 
\usepackage[T1]{fontenc}    
\usepackage[hidelinks]{hyperref}       
\usepackage{url}            
\usepackage{booktabs}       
\usepackage{amsfonts}       
\usepackage{nicefrac}       
\usepackage{microtype}      
\usepackage{url}
\usepackage{float}
\usepackage{doi}
\usepackage{subfigure}
\usepackage{multirow}
\usepackage{amsmath}
\usepackage{graphicx}
\usepackage{setspace}
\usepackage{amssymb} 
\usepackage{xcolor} 
\usepackage{subcaption}
\usepackage{xcolor,colortbl}

\usepackage[mathlines]{lineno}
\usepackage{appendix}
\usepackage{makecell} 

\definecolor{lavender}{rgb}{0.9, 0.9, 0.98}

\title{A computational framework for quantifying route diversification in road networks}

\date{}

\author{
        Giuliano Cornacchia\textsuperscript{1,2},
        Luca Pappalardo\textsuperscript{1, 3},
        Mirco Nanni\textsuperscript{1},
        Dino Pedreschi\textsuperscript{2},
        Marta C. Gonz\'{a}lez\textsuperscript{4, 5, 6} \\
    \\
    \textsuperscript{1} ISTI-CNR, Pisa, Italy \\
    \textsuperscript{2} University of Pisa, Pisa, Italy \\
    \textsuperscript{3} Scuola Normale Superiore of Pisa, Pisa, Italy \\
    \textsuperscript{4} Department of City and Regional Planning, University of California, Berkeley, CA, USA \\
    \textsuperscript{5} Energy Technologies Area, Lawrence Berkeley National Laboratory, Berkeley, CA, USA \\
    \textsuperscript{6} Department of Civil and Environmental Engineering, University of California, Berkeley, CA, USA
}

\begin{document}
\maketitle
\begin{abstract}
The structure of road networks impacts various urban dynamics, from traffic congestion to environmental sustainability and access to essential services. 
Recent studies reveal that most roads are underutilized, faster alternative routes are often overlooked, and traffic is typically concentrated on a few corridors.
In this article, we examine how road network structure, and in particular the presence of mobility attractors (e.g., highways and ring roads), shapes the counterpart to traffic concentration: route diversification. 
To this end, we introduce DiverCity, a measure that quantifies the extent to which traffic can potentially be distributed across multiple, loosely overlapping near-shortest routes.  
Analyzing 56 diverse global cities, we find that DiverCity is influenced by network characteristics and is associated with traffic efficiency. 
Within cities, DiverCity increases with distance from the city center before stabilizing in the periphery, but declines in the proximity of mobility attractors. 
We demonstrate that strategic speed limit adjustments on mobility attractors can increase DiverCity while preserving travel efficiency.
We isolate the complex interplay between mobility attractors and DiverCity through simulations in a controlled setting, confirming the patterns observed in real-world cities. 
DiverCity provides a practical tool for urban planners and policymakers to optimize road network design and balance route diversification, efficiency, and sustainability.
We provide an interactive platform ({\color{blue}\href{https://divercitymaps.github.io}{https://divercitymaps.github.io}}) to visualize the spatial distribution of DiverCity across all considered cities. 
\end{abstract}

\vfill




\onehalfspacing

\section*{Introduction}
\label{intro}

The structure of road networks profoundly influences essential aspects of urban life, such as traffic congestion \cite{saberi2020simple, Colak2016-uw, zhang2019scale}, environmental sustainability \cite{xie2007measuring, bohm2022gross, xu2016review, tenkanen2023advances}, land use patterns \cite{zeng2020effect}, the spatial organization of cities \cite{barthelemy2011spatial, barthelemy2024review}, and access to essential services and amenities \cite{tsigdinos2022rethinking, bruno2024universal}.
Recent findings reveal substantial inefficiencies in how drivers use road networks: an estimated 98\% of roads are underutilized \cite{wang2012understanding, toole2015path}, 16\% of highway-based routes have quicker alternatives \cite{wang2014empirical}, and traffic is typically concentrated into a limited number of corridors \cite{chen2024scaling, zeng2016prediction, barth2008real, Colak2016-uw}.
The growing reliance on GPS-based navigation services like TomTom and Google Maps may further amplify these inefficiencies, as both anecdotal evidence \cite{macfarlane2019navigation, Siuhi2016-gb, schade2024traffic} and recent studies suggest \cite{cornacchia2024navigation, Cornacchia2022-ni, pappalardo2023future, pedreschi2024humanai, pappalardo2024survey}.

Traffic concentration arises from the interplay between the structure of the road network, which can either support or hinder efficient traffic distribution, and the dynamics of human mobility, shaped by population patterns and travel demand. 
In this article, we investigate the role of network structure -- and particularly the presence of mobility attractors such as highways and ring roads, which are typically preferred by drivers -- in shaping the dual of traffic concentration: route diversification.
By potential route diversification, we mean the extent to which traffic can be distributed across multiple, loosely overlapping routes.
How do road networks vary in their capacity for route diversification, and how can these differences be measured? Which factors drive or hinder route diversification, and what role do mobility attractors play? Can route diversification in road networks be effectively guided through non-disruptive policy interventions? 

A review of the literature reveals a critical gap: no existing measure quantifies potential route diversification in road networks, leaving these fundamental questions unanswered.
Existing metrics fall into two categories -- edge-level and route-level -- but both have limitations.
Edge-level metrics offer insight into network flow at the level of individual road segments but cannot generalize to entire routes. For example, betweenness centrality quantifies how many shortest routes pass through an edge, ignoring the range of near-shortest alternatives that drivers may realistically take \cite{freeman1977set}. 
$K_{\text{road}}$ quantifies the contribution of different city areas to edge-level traffic but cannot assess the diversity of routes available between an origin and a destination \cite{wang2012understanding}.
Route-level metrics, in contrast, assess entire paths but still fail to capture route diversification adequately. 
The detour index measures how much a route deviates from its straight-line distance but does not account for alternative routes between the origin and destination \cite{levinson2009minimum, lee2023exploring, wang2024much}. 
The inness measure evaluates whether routes tend to converge toward or diverge from the city center and focuses only on the shortest and fastest paths, overlooking route diversity \cite{lee2017morphology}.  
In metro systems, route redundancy measures how many paths are available that are not much longer than the shortest one \cite{jing2019route}, but it fails to capture how much these alternative routes overlap in space.

We fill this gap by introducing DiverCity, a measure that quantifies both the number of practical alternative routes -- those only marginally longer than the fastest path -- and their spatial overlap.
Our measure relies exclusively on road network data, making it applicable to any city and enabling a direct assessment of how network structure influences potential route diversification. 
To ensure independence from actual travel demand and guarantee consistency across different cities, we employ the radial sampling method \cite{lee2017morphology} that generates origin-destination pairs at varying distances from the city center.
We then apply DiverCity to a dataset of 56 cities worldwide, spanning all six inhabited continents, varying population densities, and diverse road network structures.

We find substantial variability in potential route diversification across cities. 
Tokyo exhibits the highest DiverCity, while Mumbai ranks lowest. In between, grid-structured cities such as Chicago and New York tend to show higher diversification than more irregularly structured cities like Brussels and Istanbul. 
DiverCity reflects both the structural and functional properties of road networks. Indeed, it tends to be higher in cities with networks that are extensive, well-connected, and with more direct paths. Moreover, lower DiverCity is associated with trips that rely more heavily on congested roads -- where peak-hour travel times far exceed free-flow conditions.

The spatial analysis of DiverCity within cities reveals a universal pattern: it increases with distance from the city center and plateaus in peripheral areas. Mobility attractors, such as highways and ring roads, play a pivotal role in shaping this distribution. 
Indeed, DiverCity consistently decreases near these attractors as they funnel traffic into fast corridors, limiting the use of alternative routes, thereby reducing diversification for trips in their immediate vicinity. 
However, our results reveal that when strategically spaced and well-distributed, mobility attractors may enhance DiverCity at a city-wide scale. 
For example, in Rome, mobility attractors are poorly distributed, and they reduce potential route diversification considerably; in Tokyo, mobility attractors are dense and well-distributed, minimizing local suppressive effects and stabilizing DiverCity across the city. 

We propose speed limit tuning as a targeted intervention to counteract the observed suppressive effects of mobility attractors. 
We demonstrate that moderate speed reductions mitigate attractors' dominance, increasing potential route diversification considerably with only a minimal impact on travel times.
To model the impact of mobility attractors on DiverCity and uncover the mechanisms that govern it, we develop a controlled simulation on a simplified, grid-structured road network. 
Within this framework, we vary the spatial distribution and speed limits of mobility attractors to analyze their impact on DiverCity.
Our results closely align with real-world observations, confirming both the local and global effects of mobility attractors and demonstrating the effectiveness of speed limit reductions in mitigating their impact on potential route diversification.

Our study equips city planners, policymakers, and transportation authorities with valuable tools to promote more efficient use of the road network. 
DiverCity can guide infrastructure investments by pinpointing areas that would benefit from new road connections or speed limit adjustments. 
Our measure can also aid in preliminary impact assessments of urban policies like the 30 km/h speed limit policy \cite{yannis2024review, tettamanti2024relationship} by identifying how potential route diversification would respond to changes in speed regulations.
As a further contribution, we provide an interactive online platform ({\color{blue}\href{https://divercitymaps.github.io}{https://divercitymaps.github.io}}) to visualize the spatial distribution of DiverCity across all considered cities.

\section*{Measuring potential route diversification}
\label{measure}

We quantify a trip’s potential route diversification by examining the geographical characteristics of the alternative routes connecting its origin and destination.
Mathematically, we define the DiverCity of a trip $(u, v)$ as:
\begin{equation}
\mathcal{D}(u, v) = \mathcal{S}(\text{NSR}(u, v)) \cdot |\text{NSR}(u, v)| 
\label{eq:DiverCity}
\end{equation}
\noindent where $\text{NSR}(u, v)$ is the set of near-shortest routes between locations $u$ and $v$, $\mathcal{S}(\text{NSR}(u, v))$ represents their spatial spread, and $|\text{NSR}(u, v)|$ indicates the number of near-shortest routes between $u$ and $v$. 
By near-shortest routes, we refer to alternative routes whose cost deviates by up to $\epsilon$=30\% from that of the fastest route. These represent the practical route alternatives that drivers are most likely to consider when traveling to their destination.
To identify near-shortest routes, we generate up to $k$ alternative routes for a trip using path penalization \cite{cheng2019shortest, li2022comparing, hacker2021most, cornacchia2023one, cornacchia2024polaris} (see Methods for details). 
We set $k=10$ based on empirical evidence indicating that drivers' route choices are typically limited to 10 options \cite{lima2016understanding}. 
In Supplementary Note 1, we demonstrate that the results presented in this study remain robust for $k \in [2, 15]$ and $\epsilon \in \{10\%, 20\%, \dots, 50\%\}$.

Regarding the spatial spread, $\mathcal{S}(\text{NSR}(u, v))=1-J(\text{NSR}(u, v))$, where $J$ is the average weighted Jaccard similarity among all pairs of routes in $\text{NSR}(u, v)$ (see Methods for details).
$\mathcal{S}(\text{NSR}(u, v))$ captures the geographical diversity of near-shortest routes between $u$ and $v$, relying on empirical evidence indicating that drivers' route choices are typically constrained within well-defined spatial boundaries \cite{lima2016understanding}.
A high $\mathcal{S}(\text{NSR}(u, v))$ indicates minimal overlap among the routes. 
A low $\mathcal{S}(\text{NSR}(u, v))$ suggests that the near-shortest routes are highly similar, deviating only slightly from the fastest route.

$\mathcal{D}(u, v)$ ranges in $[0, k]$, where a trip with only one route has $\mathcal{D}(u, v) = 0$ and a trip with $k$ disjoint near-shortest routes has $\mathcal{D}(u, v) = k$. 
Note that trip DiverCity is not symmetric, i.e., $\mathcal{D}(u, v) \neq \mathcal{D}(v, u)$,
because of the possible existence of one-way streets between the two locations.
Figure \ref{fig:panel_1}a and \ref{fig:panel_1}b compare two trips of equal origin-to-destination distance ($\approx$ 24 km) in Mumbai and Tokyo.
The trip in Mumbai has six near-shortest routes (in blue) alongside four excessively long alternatives (in red), see Figure \ref{fig:panel_1}a.
The concentration of routes leads to substantial overlap, yielding a low $\mathcal{D}_{\text{\tiny Mumbai}}(u, v)=2.18$.
In contrast, the trip in Tokyo has many near-shortest routes with a low spatial overlap, resulting in a high 
$\mathcal{D}_{\text{\tiny Tokyo}}(u, v)=9$ (see Figure \ref{fig:panel_1}b). 

We analyze 56 cities worldwide by downloading publicly available road networks from OpenStreetMap \cite{OpenStreetMap}, each covering a 30 km radius from the city center.
The cities were selected based on their size and global relevance, comprising a mix of capital cities, major economic hubs, and megacities across multiple continents. This selection captures a broad spectrum of network structures, including planned grid-like cities (e.g., in the United States) and organically grown historical cities (e.g., in Europe). A detailed list of the cities and their characteristics is provided in Table \ref{tab:divercity_single_cities_full}.
For each city, we create a set of trips $T$ using the radial sampling method \cite{lee2017morphology}, which draws concentric circles at various distances from the city center.
The endpoints of trips in $T$ (sampled nodes) correspond to intersections in the road network.
This approach ensures that each trip’s origin and destination lie on the same circle (see Methods for details). 
Previous studies show that the radial sampling method effectively captures real mobility flow patterns in urban environments \cite{lee2017morphology, lee2023exploring, cornacchia2025path}.
By employing radial sampling, our analysis depends solely on road network information, eliminating the need for real mobility data and avoiding biases from real origin-destination matrices.
This yields a purely topological perspective on potential route diversification.

For each city $C$, we define the city-level DiverCity as the median trip DiverCity across trips in $T$, expressed as $\mathcal{D}_C = \text{median}\{\mathcal{D}(u, v) \,|\, (u, v) \in T\}$.

\section*{Results}
\begin{figure}
    \centering
    \includegraphics[width=\textwidth]{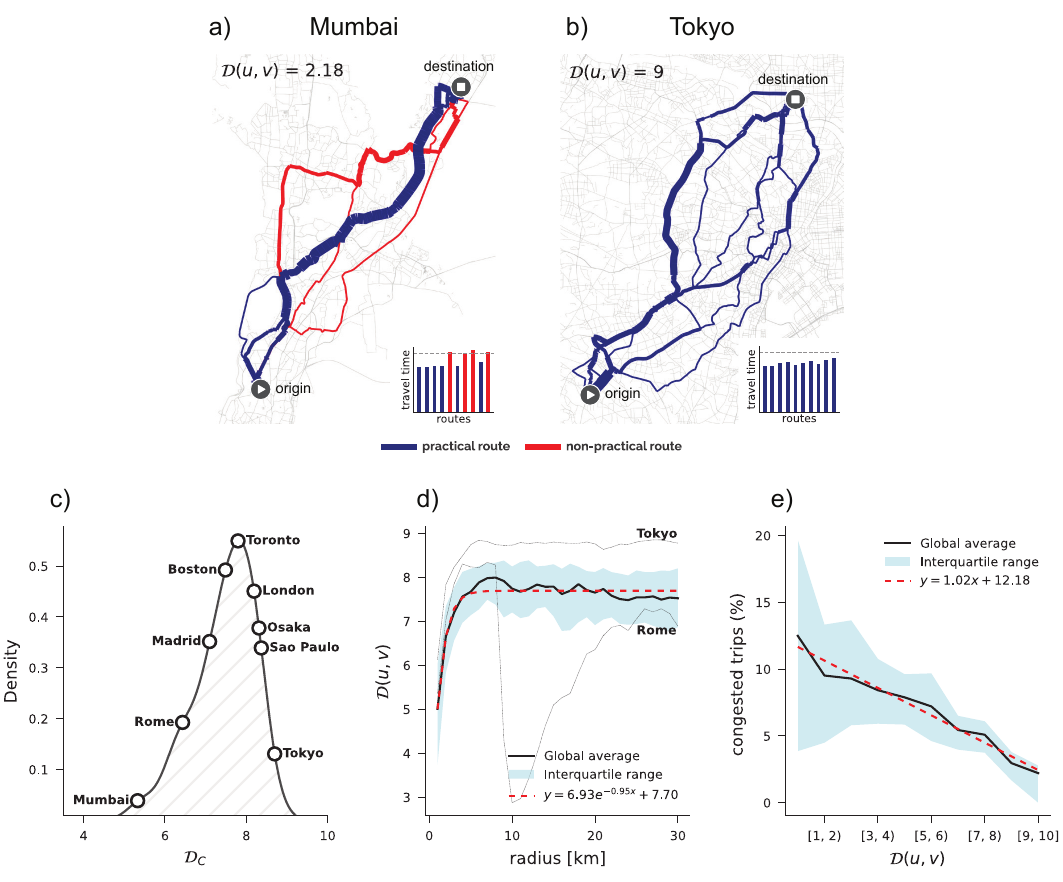}
    \caption{\textbf{Overview of DiverCity and global patterns in urban road networks.} {\textbf{(a)} Example of a trip with low DiverCity (2.18) in Mumbai. Near-shortest routes significantly overlap, leading to low route diversity.
    \textbf{(b)} Example of a trip with high DiverCity (9) in Tokyo, characterized by multiple spatially diverse near-shortest routes. For panels (a) and (b), inset bar plots show the travel time of each alternative route, with NSRs in blue and non-feasible routes (exceeding the near-shortest threshold, shown as a dashed line) in red. \textbf{(c)} Distribution of median DiverCity across 56 cities, highlighting substantial variability. \textbf{(d)} DiverCity for trips at varying radial distances from the city center. The black line shows the global average, the light blue area represents the interquartile range, and the red line is an exponential fit ($R^2 = 0.93$). Deviations include Tokyo’s high values and Rome’s localized drop near its ring road. \textbf{(e)} Percentage of congested trips at different DiverCity values. The red line shows a linear fit ($R^2 = 0.96$), and the shaded area denotes the interquartile range across cities.}}
    \label{fig:panel_1}
\end{figure}

Across the 56 cities, $\mathcal{D}_C$ exhibits a peaked distribution with a mean of $7.45 \pm 0.73$ (see Figure \ref{fig:panel_1}c). 
Tokyo ranks as the city with the highest value ($\mathcal{D}_{\text{Tokyo}} = 8.697$), while Mumbai has the lowest ($\mathcal{D}_{\text{Mumbai}} = 5.328$). 
Between these extremes, cities display varying levels of $\mathcal{D}_C$ (see Supplementary Note 2).
For example, Rio de Janeiro, Rome, and Madrid exhibit low DiverCity (5.620, 6.434, and 7.097); London, Chicago, and São Paulo exhibit high DiverCity values (8.191, 8.283, and 8.369). 
Gridded cities, comprising approximately 40\% of the sample (see Table \ref{tab:divercity_single_cities_full}), have higher and less variable $\mathcal{D}_C$ scores (7.68 $\pm$ 0.55) compared to non-gridded cities (7.29 $\pm$ 0.79). This is because gridded cities have near-shortest routes with more similar travel times than non-gridded counterparts (see Supplementary Note 3).

The spatial analysis of $\mathcal{D}(u, v)$ reveals that it is not equally distributed within a city: it is low in the city center, with an average value of 5 within the first kilometer (Figure \ref{fig:panel_1}d), and increases sharply with distance from the center, stabilizing at around 7.5 beyond 10 km.
This pattern reflects a rapid expansion of routing options in the near periphery, followed by a plateau. 
This trend is well captured ($R^2=0.93$) by a bounded exponential function: 
$\text{$y = \alpha \cdot e^{-\beta x} + \gamma$}$,
where the parameter values are $\alpha=-6.93$, $\beta=-0.95$, and $\gamma=7.70$ (red dashed curve in Figure \ref{fig:panel_1}d). 
An intriguing outlier deviates from the overall trend: in Rome, $\mathcal{D}(u, v)$ has a marked decline between 10 and 12 km from the city center (see Figure \ref{fig:panel_1}d), coinciding with the location of Rome's major ring road.

\begin{figure}
    \centering
    \includegraphics[width=\textwidth]{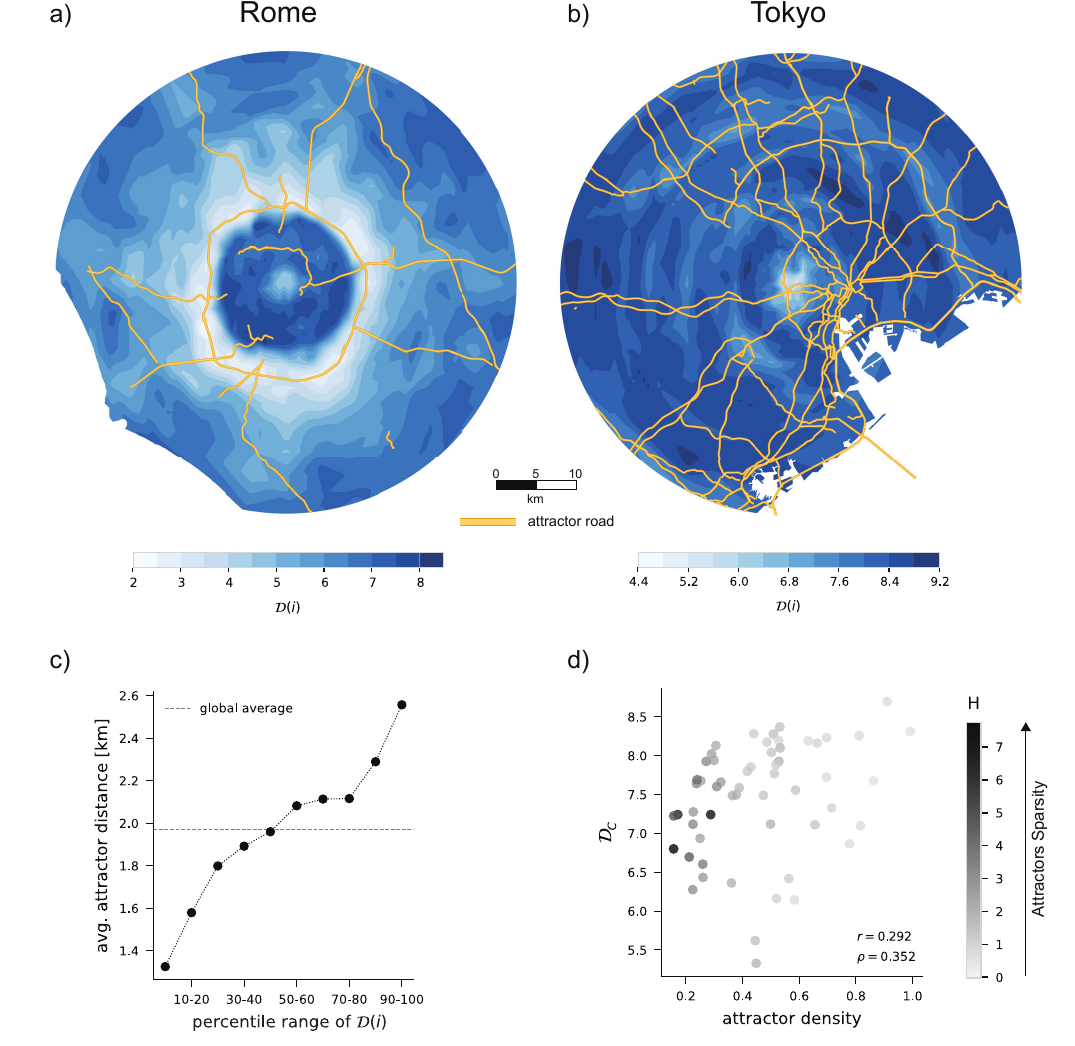}
    \caption{\textbf{DiverCity and mobility attractors.} {\textbf{(a, b)} Spatial distribution of node-level DiverCity, $\mathcal{D}(i)$, in Rome (a) and Tokyo (b). Values are interpolated between nodes, with mobility attractors highlighted in orange. Low-$\mathcal{D}(i)$ areas are represented in light blue, while high-$\mathcal{D}(i)$ areas are shown in dark blue, following the color gradient in the scale. Rome exhibits generally lower $\mathcal{D}(i)$ values and sparser attractors compared to Tokyo. In both cities, areas with low $\mathcal{D}(i)$ (relative to the city’s distribution) tend to cluster around mobility attractors. In Rome, low $\mathcal{D}(i)$ areas are strongly concentrated near the city's major ring road, while in Tokyo, they are distributed around branches of nearby mobility attractors, though less prominently than in Rome. \textbf{(c)} The average distance to the nearest attractor for nodes in different percentile ranges of $\mathcal{D}(i)$ within their respective cities. Nodes with lower $\mathcal{D}(i)$ are consistently closer to attractors. The dashed line represents the global average distance across all cities. \textbf{(d)} The relationship between city-level DiverCity ($\mathcal{D}_C$) and attractor density. Each point corresponds to a city, with color intensity indicating attractor spatial dispersion ($H$). Cities with denser and more evenly distributed attractors tend to have higher $\mathcal{D}_C$.}}
    \label{fig:panel_2}
\end{figure}

This extreme case motivated us to investigate whether reductions in $\mathcal{D}(u, v)$ generally occur near major mobility attractors across the cities under scrutiny. 
Here, we define mobility attractors as high-capacity transport infrastructures designed to facilitate and accelerate large traffic volumes, including highways, ring roads, and major arterial roads (see Methods for details). Drivers predominantly prefer mobility attractors over secondary roads \cite{thomas2015route, ramming2001network}, making it crucial to understand their impact on route diversification.
For each sampled node $i$, we measure its distance to the nearest mobility attractor and calculate its node-level DiverCity as 
$\mathcal{D}(i) = \frac{1}{2(|N|-1)} \sum_{j \in N} \big (\mathcal{D}(i, j) + \mathcal{D}(j, i) \big)$, i.e., the average trip DiverCity between $i$ and all other sampled nodes $j$ at the same radial distance ($N$).
Figure \ref{fig:panel_2} visualizes the spatial distribution of $\mathcal{D}(i)$ for two contrasting cities, Rome and Tokyo. 
In Rome (Figure \ref{fig:panel_2}a), low $\mathcal{D}(i)$ values (white and light blue areas) are strongly concentrated around Rome's ring road, which absorbs many routes and decreases route diversification within a 10–12 km radial band from the city center. 
In Tokyo (Figure \ref{fig:panel_2}b), low $\mathcal{D}(i)$ values are scattered and typically associated with specific branches of major attractor roads rather than a single dominant feature.
This distribution facilitates higher levels of route diversification throughout the city.

To isolate the impact of mobility attractors on DiverCity, we analyze nodes located more than 2 km from the city center. 
We exclude nodes within 2 km because $\mathcal{D}(i)$ is consistently low in this range across all cities (as evident from Figure \ref{fig:panel_1}d), making it difficult to distinguish the impact of mobility attractors from the inherent structural constraints of central areas.
We find a universal trend across all cities: on average, nodes with city-relative low $\mathcal{D}(i)$ -- that is, in the lower percentile ranges of their city's $\mathcal{D}(i)$ distribution -- consistently cluster nearby mobility attractors. 
For instance, nodes in the bottom 10\% of $\mathcal{D}(i)$ values are, on average, 1.33 km from the nearest mobility attractor, significantly closer than the global average distance of 2 km and the 2.56 km observed for nodes in the top 10\% of $\mathcal{D}(i)$ values (Figure \ref{fig:panel_2}c).
Mobility attractors funnel traffic into a few paths, reducing potential route diversification for trips originating or ending nearby. This effect weakens with distance, as sampled nodes farther from mobility attractors show progressively higher $\mathcal{D}(i)$ values (see Figure \ref{fig:panel_2}c).

Our results also reveal that $\mathcal{D}_C$ correlates with key road network features, including total road length ($r=0.561$, $\rho=0.652$), the number of road intersections ($r=0.431$, $\rho=0.488$), and edge circuity, i.e., the extra distance relative to the straight line path ($r=-0.297$, $\rho=-0.415$).   
See Supplementary Note 4 for a detailed analysis of 
$\mathcal{D}_C$'s relationship with road network characteristics and Table S1 for city-specific values.
Notably, cities with higher $\mathcal{D}_C$ have a higher density of attractors and more evenly distributed attractors compared to cities with lower $\mathcal{D}_C$ (see Figure \ref{fig:panel_2}d).
We measure the density of attractors in a city as their total length per km$^2$ and their spatial distribution using a dispersion index, $H$, computed as the average distance of a set of random points to the nearest mobility attractor (see Methods for details).
For instance, Tokyo, which has the highest $\mathcal{D}_C$, features half the dispersion of attractors and double their density compared to Mumbai, which ranks as the lowest in $\mathcal{D}_C$ (Table \ref{tab:divercity_single_cities_full}).
We present the DiverCity profiles and the spatial distribution of $\mathcal{D}(i)$ for all the cities in Supplementary Note 7 (Figure S24-S79).

A fundamental question is whether a trip’s DiverCity is associated with its level of congestion. We hypothesize that trips with higher DiverCity provide greater flexibility to distribute traffic across multiple alternative paths, thereby mitigating congestion.
To analyse the link between DiverCity and traffic congestion, we collect real-world travel time data from TomTom for a representative sample of trips in each city (see Methods and Supplementary Note 5).
For each trip $(u, v)$, we compute a congestion index ($CI$) as the difference between peak-hour ($t_{\text {peak}}$) and off-peak ($t_{\text {off-peak}}$) travel times, normalized by the straight-line distance, $dist(u, v)$, between origin and destination:
\begin{equation}
 CI(u, v) = \frac{t_{\text {peak}}-t_{\text {off-peak}}}{dist(u, v)}
\end{equation}
Higher $CI$ values indicate more severe congestion.

DiverCity is not correlated with $CI$ under normal traffic conditions, but a clear relationship emerges under severe congestion. Specifically, among the top 5\% most congested trips -- those with the highest $CI$ values, which we define as ``congested trips'' -- DiverCity is negatively correlated with congestion ($r = -0.227$, $\rho = -0.218$; see Supplementary Figure S12 and Note 5c). This suggests that route diversification matters most when traffic is severe.

To further explore this relationship, we compute the percentage of congested trips at different DiverCity values (see Methods for details).
We find a strong linear decrease: the percentage of congested trips drops from 12.5\% for DiverCity values near 0 to just 2.2\% near 10 (See Figure \ref{fig:panel_1}e). 
Trips with limited and spatially overlapping routing alternatives are significantly more likely to experience congestion.
This pattern is reinforced by differences in average DiverCity: congested trips have a lower mean DiverCity (5.93) than uncongested ones (6.97). At the city level, DiverCity is on average 17.9\% lower for congested trips, with this difference statistically significant in most of the cities (see Supplementary Note 5d). Importantly, this effect persists even when controlling for the use of mobility attractors: when trips are grouped by the extent to which their fastest path overlaps with attractor roads, congested trips consistently show lower DiverCity than uncongested ones across all levels of attractor reliance (see Supplementary Note 5e).

Together, these findings confirm our hypothesis that trips with higher DiverCity provide greater flexibility to distribute traffic across multiple alternative paths, thereby mitigating congestion.

An illustrative example is shown in Supplementary Figure S15, where two trips in London with identical origin-destination distance ($\approx$ 5 km) exhibit large differences in both DiverCity and $CI$. The trip with lower DiverCity experiences substantially more congestion.

\section*{Speed limits tuning on mobility attractors}

Mobility attractors are typically preferred by drivers because of their speed limits compared to other roads \cite{thomas2015route}, absorbing a high volume of routes. This accelerates traffic flow but also limits potential route diversification nearby.
How can we curb the dominance of mobility attractors, enhancing route diversification in their vicinity without compromising overall travel efficiency across the city?
We propose speed limit tuning as a strategic solution to address this challenge.

\begin{figure}
    \centering
    \includegraphics[width=\textwidth]{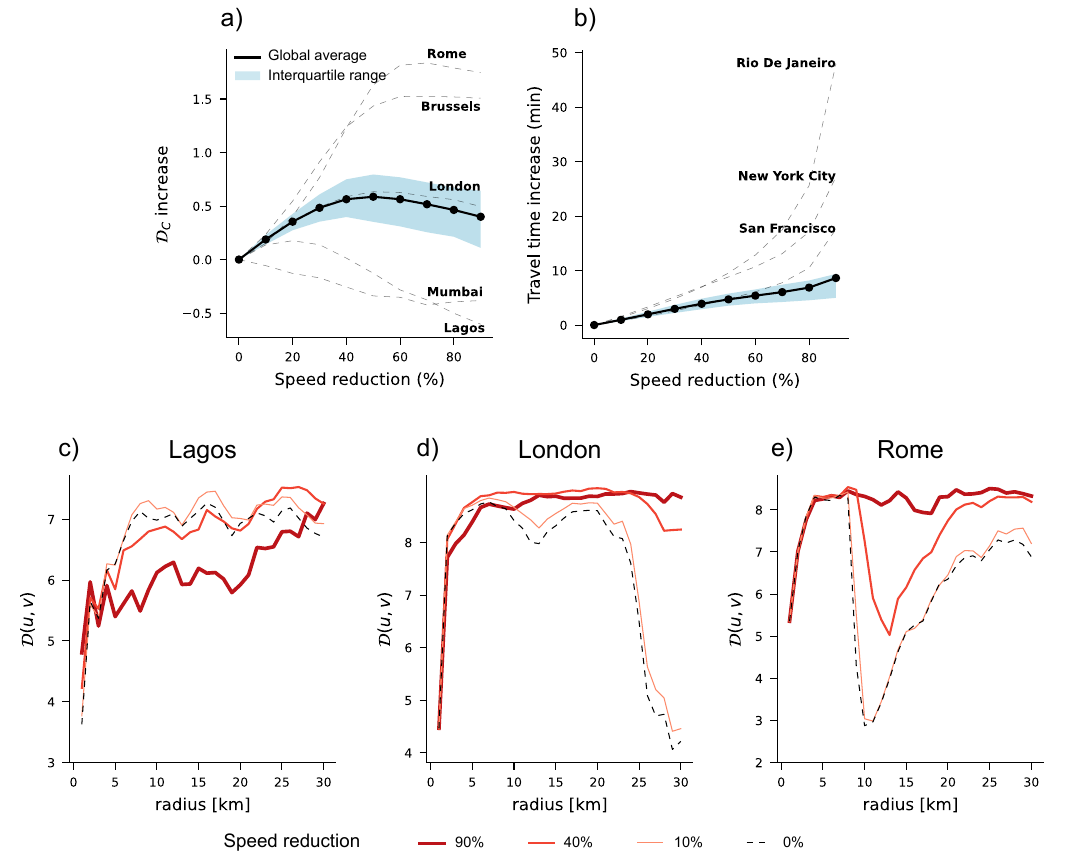}
    \caption{\textbf{Impact of speed limit tuning.} {\textbf{(a)} Effect of speed reductions on DiverCity. Cities such as Rome and Brussels show strong $\mathcal{D}_C$ improvements, while London follows the global trend with a peak at around 50\% speed reduction before stabilizing. In contrast, Mumbai and Lagos exhibit limited or negative effects. \textbf{(b)} Effect of speed reductions on travel time. Cities with critical mobility bottlenecks (e.g., bridges in Rio de Janeiro, New York City, and San Francisco) experience disproportionately large increases in travel times beyond a 50\% speed reduction. In panels (a) and (b), black lines represent the global average across all cities, while the blue shaded areas denote the interquartile range. \textbf{(c-e)} DiverCity for trips at varying radial distances for: (c) Lagos, where speed reductions negatively impact route diversification, (d) London, where DiverCity increases with speed reductions before stabilizing at 50\% as the localized effect of attractors decreases, and (e) Rome, where reductions mitigate the localized dominance of mobility attractors such as the ring road. Speed reduction scenarios (in shades of red) are compared to the baseline case (black dashed line).}}
    \label{fig:panel_3}
\end{figure}

\begin{figure}
    \centering
    \includegraphics[width=\textwidth]{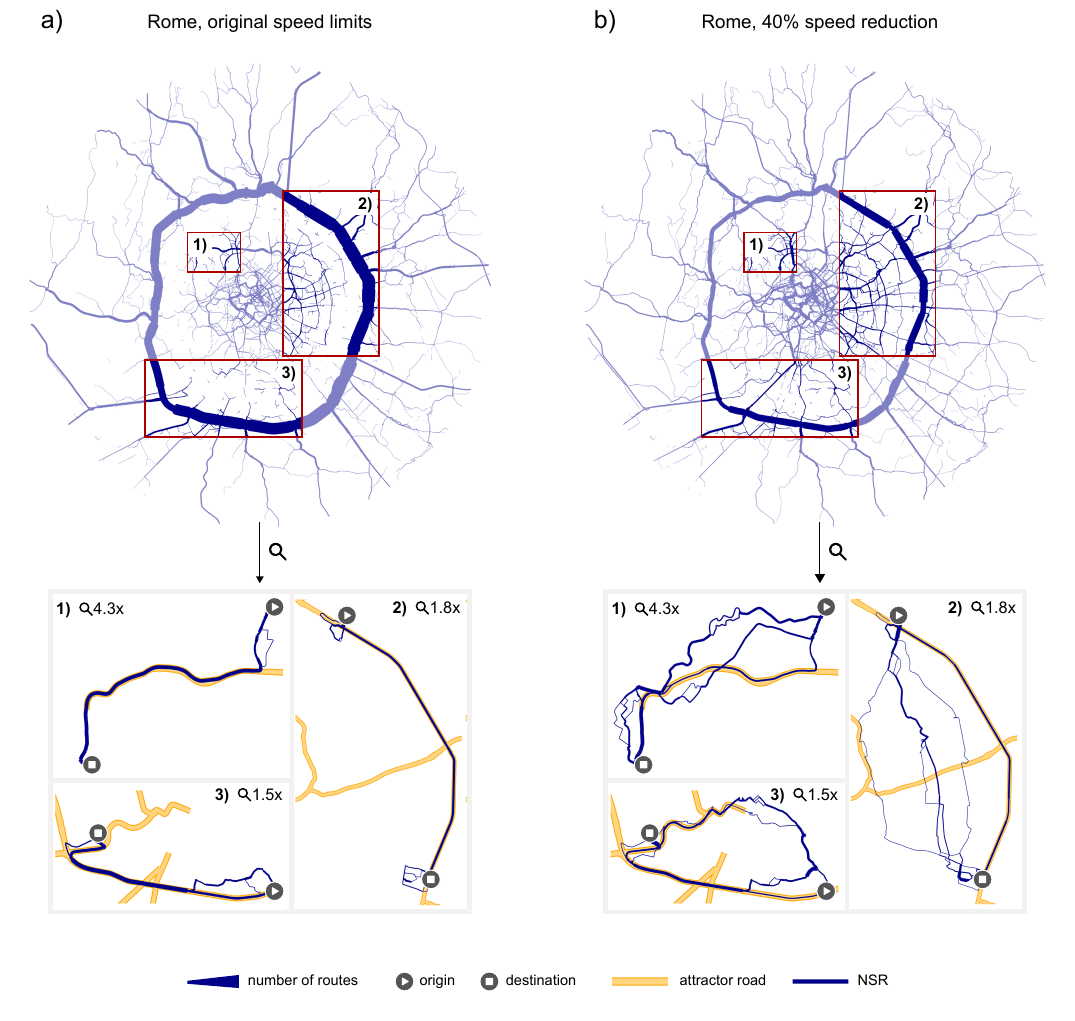}
    \caption{\textbf{Speed limit tuning in Rome.}
{Traffic distribution in Rome under original speed limits \textbf{(a)} and after a 40\% speed reduction \textbf{(b)}. The width of each road segment is proportional to the number of near-shortest routes traversing it, based on the set of sampled trips $T$. Under original speed limits, routes are highly concentrated on the ring road, suppressing potential route diversity. After a 40\% speed reduction, routes are more evenly distributed, reducing reliance on the ring road and enabling alternative routes to emerge. For each scenario, inset plots (1–3) focus on specific regions (highlighted in red on the map), providing magnified views of selected origin-destination pairs near and inside mobility attractor roads (shown in orange). The magnification factors are indicated in each inset. Under original speed limits (a), all alternative routes are funneled into the ring road, limiting route diversification. With speed reductions (b), fewer routes rely on the ring road, enabling alternative routes to emerge, increasing spatial diversity, and revealing previously "hidden" routes.}}
    \label{fig:panel_4}
\end{figure}

We simulate speed limit reductions ranging from 10\% to 90\% of original mobility attractors' values. 
In the majority of cities under study, reducing speed limits lowers the dominance of attractors and increases $\mathcal{D}_C$. 
This relationship follows a bell-shaped trend, with the largest improvements occurring at a 50\% speed reduction (Figure \ref{fig:panel_3}a). 
Beyond this threshold, there is no advantage in choosing the mobility attractors (too slow), yielding diminishing returns ($\mathcal{D}_C$ improvement decreases).
London exemplifies the typical DiverCity response to speed limit reductions: a 40\% reduction effectively eliminates the suppressive effects of mobility attractors, particularly those located around 13 km and 30 km from the city center (see Figure \ref{fig:panel_3}d and Figure S51 for a detailed visualization of attractors in London).
Certain cities exhibit distinct DiverCity responses to speed reductions (Figure \ref{fig:panel_3}a). 
For example, in Mumbai, speed reductions provide no benefits, while in Lagos, a 20\% speed reduction offers only a slight alleviation of the localized suppressive effects of mobility attractors, while further reductions lead to a consistent decrease in DiverCity improvements (Figure \ref{fig:panel_3}c). 
Rome and Brussels show instead the most significant improvements, with $\mathcal{D}{\text{\tiny Rome}}$ peaking at a 70\% speed reduction (+1.75) and $\mathcal{D}{\text{\tiny Brussels}}$ reaching its highest increase at 80\% speed reduction (+1.4).
Rome presents a particularly compelling case: the strong local impact of its ring road, located approximately 11 km from the city center, gradually weakens as speed limits are reduced (Figure \ref{fig:panel_3}e). 
As illustrated in Figure \ref{fig:panel_4}, under original speed limits, routes are heavily concentrated along the ring road, significantly suppressing route diversification (Figure \ref{fig:panel_4}a). However, implementing a 40\% speed reduction mitigates this effect, allowing alternative routes to emerge and thereby increasing $\mathcal{D}_{\text{\tiny Rome}}$ (Figure \ref{fig:panel_4}b).

While speed reductions enhance potential route diversification, they also cause a moderate increase in travel times. 
For instance, a 50\% speed reduction results in trips that are five minutes longer on average (Figure \ref{fig:panel_3}b).
A 30\% reduction yields notable benefits with minimal trade-offs: on average, $\mathcal{D}_C$ improves by 0.483 while travel times increase by two minutes only.

Bridge-dominated cities, such as San Francisco, New York City, and Rio de Janeiro, also present a distinctive pattern. 
At speed reductions of up to 40\%, these cities follow the global trend of moderate travel time increases, but beyond this threshold, speed reductions lead to disproportionately large increases in travel times. 
As shown in Figure \ref{fig:panel_3}b, at 80\% reductions, travel time increases by 10 min in San Francisco, 17 min in New York City, and 25 min in Rio de Janeiro, compared to a global average of about 7 min. 
At 90\% reductions, these values escalate further compared to the global average of around 9 min: 18 min in San Francisco, 27 min in New York City, and an extreme increase of 48 min in Rio de Janeiro.
Since bridges channel most routes between the two sides of these cities, reducing their speed limits impacts travel times for all traversing routes. Moderate reductions ($<60$-$70\%$) enhance $\mathcal{D}_C$ with minimal travel time increases, but beyond this threshold, benefits fade as travel times rise disproportionately.

\section*{Simulations in a controlled setting}
\label{toymodel}

\begin{figure}
    \centering
    \includegraphics[width=\textwidth]{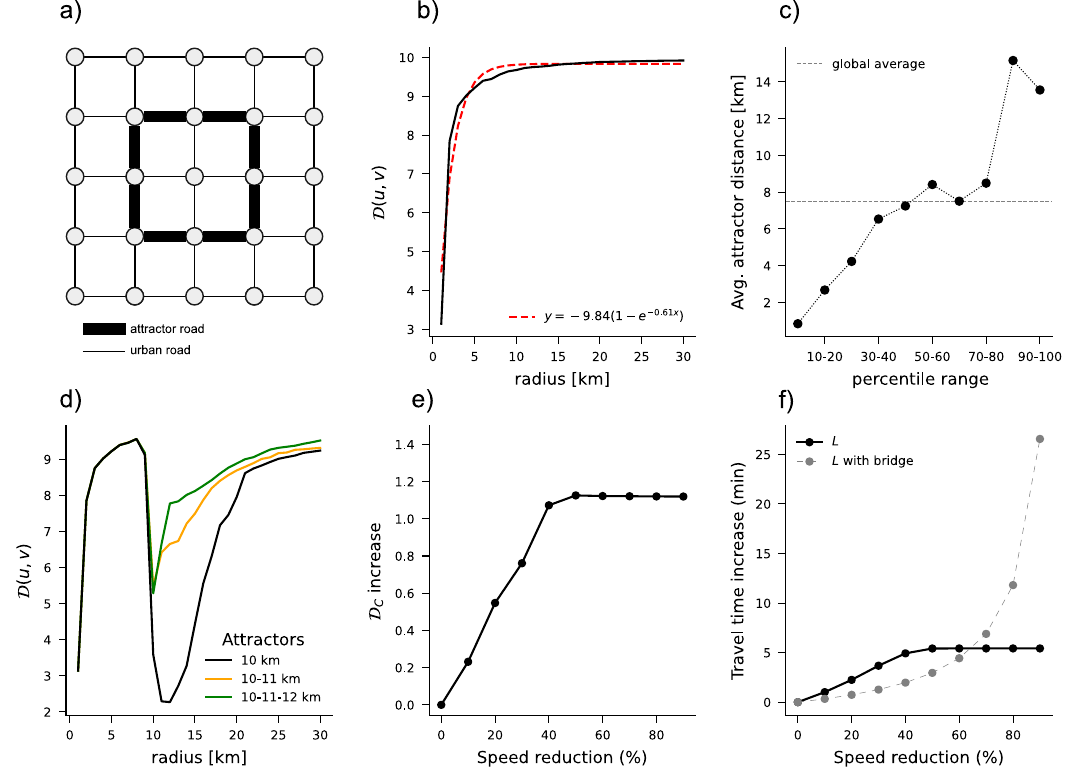}
    \caption{\textbf{Simulations in a controlled setting}. \textbf{(a)} Illustration of the lattice grid model $L$, where intersections are represented as nodes and roads as edges. Thicker edges indicate mobility attractors with higher speed limits, while the remaining edges represent standard urban roads. \textbf{(b)} DiverCity, $\mathcal{D}(u, v)$, as a function of radial distance from the center of the lattice. Without mobility attractors, $\mathcal{D}(u, v)$ increases rapidly near the center and plateaus farther out, following a bounded exponential trend (red dashed line, $R^2 = 0.93$). \textbf{(c)} The average distance to the nearest attractor for nodes grouped by percentile ranges of $\mathcal{D}(i)$ in the lattice model. Nodes with lower $\mathcal{D}(i)$ are consistently closer to attractors, mirroring the trends observed in real-world road networks. \textbf{(d)} The effect of introducing attractors at varying distances from the center on $\mathcal{D}(u, v)$. A single attractor at 10 km sharply reduces $\mathcal{D}(u, v)$ near its location. Adding more attractors at greater distances (e.g., 10–12 km) helps stabilize and increase $\mathcal{D}(u, v)$ globally. \textbf{(e)} The effect of speed limit reductions on city-level DiverCity ($\mathcal{D}_C$) in the lattice. Reducing attractor speeds increases $\mathcal{D}_C$, with the largest improvements at around 50\% speed reduction. Further reductions yield diminishing returns as attractors lose their dominance. \textbf{(f)} The impact of speed limit reductions on average travel time. For a simple lattice ($L$), travel time rises modestly with reductions. However, introducing a bridge-like bottleneck significantly amplifies travel time increases at higher speed reductions, reflecting the trend observed for real-world road networks.}
    \label{fig:panel_5}
\end{figure}

To isolate the causal effects of the placement and speed limits of mobility attractors on DiverCity in a controlled environment, we model a city as a uniform lattice $L$ of intersections (nodes) and road segments (edges), see Figure \ref{fig:panel_5}a.
$L$ spans an area of $60 \times 60$ km, matching the scale used for real cities (a 30 km radius).
Each road segment in $L$ has a length of 500 meters and a default speed limit of 50 km/h.
We introduce mobility attractors in $L$ as a ``square'' centered at $L$'s midpoint, with a side length of $2d$ and a speed limit of 100 km/h.
We generate a set of trips using radial sampling, following the same procedure applied to real cities. 
We then perform simulations by varying the position and reducing the speed limits of mobility attractors while keeping the set of trips fixed.

The simulation results confirm the findings observed for real cities.
First, when $L$ has no mobility attractors, $\mathcal{D}_L(u, v)$ increases rapidly with distance from the city center before plateauing (see Figure \ref{fig:panel_5}b).
Second, introducing a mobility attractor $A_d$ at a distance $d$ from the center of the lattice reduces $\mathcal{D}_L(u, v)$ in its vicinity, the extent of this reduction being independent of $d$ and increasing with the speed limit $A_d$ (see Supplementary Notes 6a and 6b).
Sampled nodes in the lower percentile ranges of $\mathcal{D}_L(i)$ consistently cluster near attractors, as observed in real cities. For instance, for $d=10$ km, low-$\mathcal{D}_L(i)$ nodes are, on average, just 0.85 km from the nearest attractor (Figure \ref{fig:panel_5}c). 
Third, the presence of multiple attractors stabilizes $\mathcal{D}_L(u, v)$ globally. 
Specifically, introducing a second attractor $B_{d + \delta}$ in an offset $\delta$ from $A_d$ provides limited benefits near the center but becomes more effective farther from it (see Supplementary Note 6c). 
The benefit decreases linearly as $\delta$ increases, and this effect generalizes to larger configurations: attractors clustered farther from the grid center stabilize $\mathcal{D}_L(u, v)$ more effectively, with the benefit decreasing linearly as the offset increases (see Supplementary Note 6d). 
This behaviour is exemplified in Figure \ref{fig:panel_5}d, showing $\mathcal{D}_L(u, v)$ under three configurations. 
A single attractor at 10 km from the center sharply reduces $\mathcal{D}_L(u, v)$ nearby. Adding a second attractor at 11 km from the center improves DiverCity locally, while three clustered attractors (at 10, 11, and 12 km) stabilize and enhance $\mathcal{D}_L(u, v)$ globally.

Finally, as observed in real cities, reducing attractors' speed limits in $L$ improves $\mathcal{D}_L$ linearly until the attractor speed aligns with other roads, after which improvements plateau (Figure \ref{fig:panel_5}e). However, we identify a key discrepancy: in $L$, travel times increase linearly up to this equilibrium point (50\%) and then stabilize, whereas in real cities, travel times continue to rise beyond equilibrium. 
This difference arises from the presence of bottlenecks in real cities, such as bridges, which remain critical even at reduced speed limits. To incorporate bottlenecks into $L$, we divide it into two unconnected regions connected by an attractor acting as a bridge (see Figure S23). In this modified setup, travel times continue to rise beyond equilibrium, closely mirroring the behavior of real cities (Figure \ref{fig:panel_5}f).

\section*{Discussion}
\label{dicussion}

Since mobility attractors draw traffic towards themselves, their spatial distribution affects potential route diversification in their vicinity.  
Rome and Tokyo represent two contrasting examples of this phenomenon: Rome, with its single dominant ring road, exhibits low DiverCity; Tokyo, with its dense and widespread network of attractors, shows high DiverCity.
Mobility attractors favour fluid movement, but they also reduce route diversity to protect local streets, which raises a critical question: is high DiverCity always desirable?
Our analysis reveals a nuanced trade-off: higher DiverCity is consistently associated with trips that avoid heavily congested roads, where peak-hour travel times substantially exceed free-flow conditions. This makes DiverCity a valuable proxy for assessing road network performance. 
However, this benefit diminishes when trips rely heavily on attractors. By quantifying how infrastructure channels traffic, DiverCity offers planners a clear lens to distinguish between intended design outcomes and unintended inefficiencies, enabling more balanced, resilient urban mobility strategies.

The impact of mobility attractors on route diversification depends not only on their presence but also on how they are integrated into the urban fabric.
The distance of mobility attractors from the city center, their spacing relative to one another, their distribution across the road network, and their speed limits are all critical design factors that determine whether route diversification is enhanced or suppressed. 
Effective urban planning is not merely about constructing high-speed roads but integrating them properly into the road network to trade off traffic fluidity and potential route diversification.
DiverCity supports simulations and scenario analyses, helping identify optimal speed limit adjustments on attractors to maximize potential route diversification. 
Supplementing this analysis with empirical data -- such as vehicular GPS traces -- could shed light on the gap between a city’s potential route diversification and the actual routes drivers take on the road network. This would help quantify both the potential influence of attractors and how drivers capitalize on these opportunities in their everyday travel.

An intriguing direction emerging from our study is investigating the relationship between route diversification and urban policies such as the 30 km/h City and the 15-minute City. 
The 30 km/h City concept promotes road safety by lowering speed limits to 30 km/h on all urban streets except major roadways (i.e., mobility attractors). 
In cities where this policy has been implemented, it led to a reduction of accidents and traffic while encouraging greater use of bicycles and public transportation, with only a minor impact on travel times \cite{yannis2024review, tettamanti2024relationship}. 
However, the impact of the 30 km/h policy on potential route diversification remains uncharted territory. 
Tools like those introduced in this study could help simulate how drivers adjust their routes in response to the policy and identify which roads would be affected by their alternative routes. 
Additionally, examining the interplay between speed limit adjustments on mobility attractors and roads affected by the 30 km/h policy could help identify optimal speed limit combinations that balance policy objectives, traffic flow, and route diversification.

The 15-minute city model suggests that ensuring essential services and amenities are accessible within a 15-minute walk or bike ride can improve efficiency, equity, and sustainability \cite{moreno2021introducing, lima2023quest}. 
Recent research indicates that only a small fraction of cities worldwide meet these criteria \cite{bruno2024universal, yang2023visualizing, vale2023accessibility}.
Promoting route diversification -- through strategies like speed limit adjustments, as proposed in our work -- can enhance connectivity between neighborhoods, improve access to essential services and amenities, and serve as a practical step toward achieving the vision of a 15-minute city.

In conclusion, our study advances the understanding of vehicular routing in urban environments while illuminating its intricate ties to road network structure. Building on these insights, we introduce practical, actionable tools to adapt and design road networks, fostering diverse routes. 
At the same time, our findings highlight the dual role of mobility attractors, offering rapid connections between city areas yet potentially suppressing alternative routes. 
By providing both a deeper understanding and tangible methodologies for intervention, our research sparks further exploration and equips urban planners to shape more equitable, resilient cities.

\section*{Methods}
\label{methods}

\paragraph{Road Network Representation.}
To analyze route diversification in urban road networks, we used detailed representations of the road infrastructures in 56 global cities. Each city's road network is modeled as a directed weighted multigraph $G=(V, E)$, where $V$ denotes the set of nodes $v_i$ representing intersections, and $E$ is a multiset of edges representing the road segments connecting the vertices. Each edge $e_{i,j} \in E$ is associated with its minimum expected travel time, length, capacity, and speed limit.
To ensure consistency in our analysis, we defined the city center for each urban area using geographic coordinates referenced from {\color{blue}\href{https://www.latlong.net/}{latlong.net}}. We extracted the road network of each city from OpenStreetMap \cite{OpenStreetMap} using OSMnx \cite{boeing2017osmnx}, centering it on the city’s center and extending approximately 30 km in radius. This distance encompasses a substantial portion of the urban and peri-urban road network, allowing for a comprehensive evaluation of route diversification within a broad geographic scope.

\paragraph{Radial Sampling.}
To systematically assess potential route diversification, we employed a radial sampling approach to generate origin-destination (OD) pairs by scanning the network radially from the city center. This method, widely used to analyze spatial metrics and urban patterns within cities \cite{lee2017morphology, lee2023exploring}, ensures a uniform distribution of OD pairs across varying distances from the city center, facilitating a comprehensive evaluation of potential route diversification while avoiding sample bias.
We identified each city's center based on geographic coordinates sourced from {\color{blue}\href{https://www.latlong.net/}{latlong.net}}. From this center, we defined concentric circles with radii ranging from 1 km to 30 km at intervals of 1 km, representing increasing distances from the urban core to the periphery. The maximum radius of 30 km was chosen to sufficiently cover the urbanized areas of our sample cities \cite{lee2017morphology, lee2023exploring}. Along each circle, we selected points at 10° intervals, yielding 36 equally spaced points along the circumference. Each point was then matched to the nearest road network node within a distance threshold of 500 m. Points that could not be matched --for example, those in inaccessible areas such as water bodies, forests, or unconnected terrain-- were excluded from the sample.
In an ideal scenario, where all points are successfully matched, this process generates up to 1,260 OD pairs per radius, yielding a maximum of 37,800 OD pairs per city. However, the actual number of pairs varies based on the city's topology and the accessibility of its road network.

\paragraph{Near Shortest Routes (NSR).}
To compute the near-shortest routes (NSR) between an origin-destination (OD) pair, we perform two key steps:
\begin{enumerate}
    \item \textbf{Generating Alternative Routes}: First, we generate up to $k$ alternative routes using the Path Penalization (PP) \cite{johnson1993HIGHWAY, ruphail1995decision} algorithm, a widely used and robust alternative routing method that forms the foundation for several advanced algorithms \cite{cheng2019shortest, li2022comparing, hacker2021most, cornacchia2023one, cornacchia2024polaris}. This algorithm generates alternative routes by iteratively penalizing the weights of edges contributing to the current fastest path \cite{akgun2000finding, bader2011alternative, cheng2019shortest, li2022comparing}. The value of $k$ is set by design, but the actual number of routes returned may be fewer if no additional unique routes are found. Specifically, in each iteration, PP computes the fastest path (using Dijkstra’s algorithm) and increases the weights of its constituent edges by a factor $p$, such that $w(e) = w(e) \cdot (1 + p)$ \cite{cheng2019shortest}, where $w(e)$ denotes the expected travel time of edge $e$. Penalized edges become less likely to be chosen in subsequent iterations, prompting the exploration of alternative routes. The penalization is cumulative: edges used in earlier iterations that reappear in subsequent fastest paths receive additional penalties \cite{cheng2019shortest}.
    The higher the penalty factor $p$, the greater the deviation from the original fastest path \cite{kobitzs2013chevolution}. In our experiments, we set $p=0.1$, and we show that results are consistent across a range of $p$ values (see Supplementary Note 1).

    \item \textbf{Filtering Near Shortest Routes}: From the set of up to $k$ generated routes, we identify the NSR as the routes whose costs do not exceed the cost of the fastest route by more than $\epsilon\%$ \cite{hacker2021most}. This ensures that the resulting routes are practical alternatives that are close in cost to the optimal route. In our experiments, we set $\epsilon = 30\%$, allowing the NSRs to have costs up to 30\% higher than the fastest path. Further analyses confirm that results remain consistent across different $\epsilon$ values (see Supplementary Note 1).
    
\end{enumerate}

We compute the spatial spread $\mathcal{S}$ of the NSR between an origin-destination pair $(u, v)$ as $\mathcal{S}(\text{NSR}(u, v))=1-J(\text{NSR}(u, v))$, where $J$ is the average weighted Jaccard similarity among all pairs of routes in $\text{NSR}(u, v)$. The weighted Jaccard similarity~\cite{weighted_jaccard_2021} accounts for the length of each road segment and is defined as: $J(A, B) = \frac{\sum_{e \in A \cap B} w_e}{\sum_{e \in A \cup B} w_e}$ where $A$ and $B$ are the routes represented as sets of road segments, and $w_e$ is the length of road segment $e$. A high $\mathcal{S}(\text{NSR}(u, v))$ value indicates greater spatial spread, reflecting more diverse routing options. Conversely, a low value suggests that the routes are geographically similar and significantly overlap.

\paragraph{Mobility Attractors.}
In our study, we define mobility attractors as high-capacity transport infrastructures designed to facilitate and accelerate large traffic volumes, including highways, ring roads, and major arterial roads. To identify mobility attractors in each city, we employ a systematic approach using OpenStreetMap road network data.
Our classification of mobility attractors focused on two primary road types commonly designed to handle significant traffic volumes:
\begin{itemize}
    \item \textbf{Motorways} \texttt{(tag OSM: motorway)}: High-capacity, high-speed routes designed for fast, long-distance travel with limited access points.
    \item \textbf{Trunks} \texttt{(tag OSM: trunk)}: Major roads connecting important regions, supporting significant traffic flow just below motorway levels.
\end{itemize}
The identification process allows us to capture the physical characteristics of these roads and their intended role in traffic planning.

\paragraph{Spatial Dispersion of Mobility Attractors.}
To quantify the spatial dispersion of mobility attractors across the geographical area of interest, we employ a sampling-based methodology. We generate $N$=20,000 random points distributed uniformly within the area of interest while systematically excluding inaccessible regions such as water bodies, forests, and other non-urbanized zones. For each randomly generated point, we compute the distance $H$ to the nearest mobility attractor. The distance metric for each city is then defined as the average of these distances. Lower values indicate a more uniform and dense distribution of attractors, implying high coverage and accessibility. Conversely, higher values suggest a sparse and uneven distribution of attractors, pointing to limited coverage and diminished accessibility in specific areas.

\paragraph{Congestion Index.}
To quantify trip-level congestion, we compute a congestion index $CI$, which measures the excess travel time per unit distance under peak conditions relative to free-flow conditions. For a trip $(u,v)$, the congestion index is defined as $CI(u, v) = \frac{t_{\text {peak}}-t_{\text {off-peak}}}{dist(u, v)}$, where $t_{\text{peak}}$ and $t_{\text{off-peak}}$ are the estimated travel times during peak and off-peak conditions, respectively, and $dist(u,v)$ is the straight-line distance between $u$ and $v$. $CI$ is expressed in seconds per kilometer.
Travel time estimates are retrieved from TomTom, a widely used commercial navigation service that provides route recommendations based on real-time and historical traffic data \cite{cornacchia2024navigation}. Peak and off-peak travel times are retrieved by setting the departure time in the request to 8 AM and 2 AM local time, respectively, corresponding to rush hour and free-flow conditions.
We compute $CI$ over a representative subset of 2,142 OD pairs per city, sampled from the radial sampling at distances of 2, 4, 6, 8, 10, 12, and 15 km from the city center, with nodes spaced every 20°. We classify trips in the top 5\% of a city’s $CI$ distribution as \textit{congested}. To compute the percentage of congested trips, we first discretise the trips’ DiverCity values into integer bins from 0 to 10, the minimum and maximum DiverCity values, respectively, with a bin width of 1. For each bin, the percentage of congested trips is calculated as the number of congested trips in the bin divided by the total number of trips in the bin, multiplied by 100. This yields, for each DiverCity value, the percentage of trips experiencing congestion.

\newpage


\clearpage 

\clearpage 

\clearpage 

\clearpage 

\clearpage

\newpage
\begin{table}
{\tiny
\resizebox{\textwidth}{!}{
\begin{tabular}{l l c c c c c |@{\hspace{20pt}}| l l c c c c c}
\toprule
\textbf{rank} & \textbf{City} & \textbf{$\mathcal{D}_C$} & \textbf{attractors length} & \textbf{attractors density} & \textbf{H [km]} & \textbf{Pop. Density} &
\textbf{rank} & \textbf{City} & \textbf{$\mathcal{D}_C$} & \textbf{attractors length} & \textbf{attractors density} & \textbf{H [km]} & \textbf{Pop. Density} \\
\midrule
1 & Tokyo & 8.697 & 2,534 & 0.910 & 1.339 & 6,169 &
29 & Detroit $\boxplus$ & 7.603 & 844 & 0.309 & 3.761 & 1,750 \\
\rowcolor{lavender!80} 2 & São Paulo & 8.369 & 1,652 & 0.532 & 2.215 & 8,055 &
30 & Vancouver $\boxplus$ & 7.588 & 481 & 0.389 & 2.383 & 5,493 \\
3 & Osaka & 8.311 & 2,808 & 0.992 & 1.064 & 12,111 &
31 & Sydney & 7.557 & 1,042 & 0.587 & 2.097 & 400 \\
\rowcolor{lavender!80} 4 & Chicago $\boxplus$ & 8.283 & 799 & 0.440 & 1.769 & 4,663 &
32 & Bangkok & 7.495 & 1,155 & 0.378 & 2.642 & 5,293 \\
5 & Melbourne $\boxplus$ & 8.279 & 1,282 & 0.508 & 2.249 & 500 &
33 & San Francisco $\boxplus$ & 7.489 & 753 & 0.474 & 2.154 & 7,171 \\
\rowcolor{lavender!80} 6 & New York City $\boxplus$ & 8.257 & 2,128 & 0.811 & 1.381 & 11,316 &
34 & Boston & 7.487 & 950 & 0.364 & 2.784 & 5,200 \\
7 & Shanghai & 8.231 & 2,012 & 0.697 & 1.422 & 3,922 &
35 & Tehran & 7.328 & 1,578 & 0.714 & 1.617 & 12,028 \\
\rowcolor{lavender!80} 8 & Philadelphia $\boxplus$ & 8.197 & 1,680 & 0.528 & 1.602 & 4,129 &
36 & Hamburg & 7.277 & 683 & 0.226 & 3.380 & 2,366 \\
9 & London & 8.191 & 2,012 & 0.632 & 1.363 & 5,614 &
37 & Bogotá $\boxplus$ & 7.242 & 441 & 0.172 & 6.510 & 5,018 \\
\rowcolor{lavender!80} 10 & Los Angeles $\boxplus$ & 8.174 & 1,278 & 0.486 & 1.762 & 3,287 &
38 & Manila $\boxplus$ & 7.241 & 553 & 0.288 & 7.162 & 41,515 \\
11 & Seoul & 8.161 & 2,068 & 0.664 & 1.563 & 16,552 &
39 & Ottawa $\boxplus$ & 7.226 & 428 & 0.158 & 5.414 & 334 \\
\rowcolor{lavender!80} 12 & New Delhi & 8.130 & 960 & 0.306 & 2.836 & 11,289 &
40 & Jakarta & 7.119 & 1,225 & 0.499 & 2.634 & 15,292\\
13 & Mexico City $\boxplus$ & 8.099 & 1,580 & 0.533 & 2.520 & 6,202 &
41 & Athens & 7.117 & 425 & 0.225 & 3.854 & 17,040\\
\rowcolor{lavender!80} 14 & Houston $\boxplus$ & 8.042 & 1,574 & 0.502 & 2.110 & 1,497 &
42 & Kuala Lumpur & 7.111 & 1,620 & 0.655 & 2.100 & 7,276 \\
15 & Buenos Aires $\boxplus$ & 8.022 & 510 & 0.291 & 3.044 & 15,046 &
43 & Madrid & 7.097 & 1,833 & 0.816 & 1.359 & 5,390\\
\rowcolor{lavender!80} 16 & Milan & 7.939 & 932 & 0.299 & 2.941 & 7,700 &
44 & Brussels & 6.935 & 800 & 0.250 & 3.016 & 7,489\\
17 & Cairo & 7.928 & 1,499 & 0.528 & 2.636 & 3,256 &
45 & Shenzhen & 6.868 & 1,798 & 0.777 & 1.244 & 8,534 \\
\rowcolor{lavender!80} 18 & Lima $\boxplus$ & 7.926 & 432 & 0.272 & 3.756 & 3,329 &
46 & Kinshasa & 6.801 & 209 & 0.157 & 7.749 & 1,713 \\
19 & Dallas $\boxplus$ & 7.883 & 1,622 & 0.518 & 1.832 & 1,525 &
47 & Lagos & 6.696 & 439 & 0.212 & 5.108 & 6,871 \\
\rowcolor{lavender!80} 20 & Paris & 7.854 & 1,341 & 0.430 & 1.906 & 20,460 &
48 & Dhaka & 6.604 & 780 & 0.260 & 4.226 & 30,460\\
21 & Toronto $\boxplus$ & 7.796 & 740 & 0.417 & 2.009 & 4,336 &
49 & Rome & 6.434 & 741 & 0.261 & 3.309 & 2,232 \\
\rowcolor{lavender!80} 22 & Washington D.C. $\boxplus$ & 7.769 & 1,620 & 0.512 & 1.976 & 3,969 &
50 & Barcelona $\boxplus$ & 6.417 & 870 & 0.564 & 1.812 & 15,980 \\
23 & Beijing $\boxplus$ & 7.724 & 2,218 & 0.696 & 1.273 & 1,334 &
51 & Amsterdam & 6.362 & 898 & 0.361 & 2.642 & 5,265 \\
\rowcolor{lavender!80} 24 & Karachi & 7.689 & 416 & 0.240 & 4.733 & 26,629 &
52 & Moscow & 6.276 & 688 & 0.224 & 3.654 & 5,257\\
25 & Berlin & 7.677 & 722 & 0.253 & 3.063 & 4,227 &
53 & Istanbul & 6.161 & 996 & 0.520 & 1.649 & 2,987 \\
\rowcolor{lavender!80} 26 & Guangzhou & 7.676 & 2,749 & 0.862 & 1.035 & 2,512 &
54 & Dubai $\boxplus$ & 6.145 & 939 & 0.583 & 1.294 & 860 \\
27 & Santiago $\boxplus$ & 7.660 & 679 & 0.323 & 3.271 & 10,748 &
55 & Rio de Janeiro & 5.620 & 796 & 0.445 & 2.314 & 5,340 \\
\rowcolor{lavender!80} 28 & Cape Town & 7.641 & 322 & 0.238 & 3.842 & 1,530 &
56 & Mumbai & 5.328 & 739 & 0.448 & 2.652 & 20,694 \\
\bottomrule
\end{tabular}}}
\caption{\textbf{City-level DiverCity ranking.} {The table presents a comparative analysis of 56 cities, ranked by their city-level DiverCity ($\mathcal{D}_C$). For each city, we include the total length of mobility attractors (e.g., highways, major roads), their density (defined as attractor length per km$^2$), the spatial dispersion of attractors ($H$), and the city's population density. A $\boxplus$ symbol next to the city name indicates that the city exhibits a gridded road network structure.}}
\label{tab:divercity_single_cities_full}
\end{table}

\newpage

\nolinenumbers

\paragraph{Acknowledgements}
LP and GC have been supported by PNRR (Piano Nazionale di Ripresa e Resilienza) in the context of the research program 20224CZ5X4 PE6 PRIN 2022 "URBAI – Urban Artificial Intelligence" (CUP B53D23012770006), funded by the European Commission under the Next Generation EU programme. 
DP has been supported by PNRR - M4C2 - Investimento 1.3, Partenariato Esteso PE00000013 - ”FAIR – Future Artificial Intelligence Research” – Spoke 1 "Human-centered AI", funded by the European Commission under the NextGeneration EU programme.
MN has been partially supported by Project "SoBigData.it - Strengthening the Italian RI for Social Mining and Big Data Analytics", prot. IR0000013, avviso n. 3264 on 28/12/2021.
\\We thank Daniele Fadda for his precious support with data visualization and Giovanni Mauro, Minho Kim, Margherita Lalli, Antonio Desiderio, Daniele Gambetta, Gabriele Barlacchi, and Charlotte Brimont for their useful suggestions.

\paragraph{Data availability statement}
We compute the DiverCity measure using solely road network data from OpenStreetMap \cite{OpenStreetMap}, extracted with OSMnx \cite{boeing2017osmnx}. For the congestion index analysis, we used routing suggestions from TomTom’s APIs, which cannot be shared due to proprietary restrictions.
The Python code used to download the road networks for the cities analyzed in this work is publicly available at {\color{blue}\href{https://github.com/GiulianoCornacchia/DiverCity}{https://github.com/GiulianoCornacchia/DiverCity}}.

\paragraph{Code availability statement}
The Python code required to fully reproduce the analyses presented in this study is publicly available at {\color{blue}\href{https://github.com/GiulianoCornacchia/DiverCity}{https://github.com/GiulianoCornacchia/DiverCity}}.

\paragraph{Author contributions} 
GC conceptualized the study, developed the code of measures and experiments, conducted the experiments, designed and created the figures, developed the online platform, and wrote the manuscript. 
LP conceptualized the study, designed the experiments, designed the figures, wrote the manuscript, and supervised the research. 
MN and DP conceptualized the study and critically revised the manuscript. 
MG conceptualized the study, designed the experiments, revised the manuscript, and supervised the research. 
All authors reviewed and approved the final version of the manuscript.

\paragraph{Competing interests} The authors declare no competing interests.

\bibliographystyle{abbrv} 
\bibliography{biblio}  

\end{document}